\documentstyle[12pt,fleqn,epsfig]{article}

\begin{document}

\renewcommand{\topfraction}{0.99}
\renewcommand{\bottomfraction}{0.99}
\renewcommand{\textfraction}{0.01}
\setcounter{totalnumber}{6}

\begin{center}
  {\Large \bf Search for Deconfinement in NA49 at the CERN SPS } 
\end{center}

\vspace{0.5cm}
\begin{center}
Peter Seyboth, Max-Planck-Institut f\"ur Physik, 80805 Munich, Germany
\end{center}

\vspace{0.3cm}
\begin{center}
for the NA49 collaboration:
\end{center}

\vspace{0.3cm}
\noindent{
S.V.~Afanasiev$^{9}$, T.~Anticic$^{19}$, D.~Barna$^{5}$,
J.~Bartke$^{7}$, R.A.~Barton$^{3}$,
L.~Betev$^{10}$, H.~Bia{\l}\-kowska$^{17}$, A.~Billmeier$^{10}$,
C.~Blume$^{8}$, C.O.~Blyth$^{3}$, B.~Boimska$^{17}$, M.~Botje$^{1}$,
J.~Bracinik$^{4}$, R.~Bramm$^{10}$, R.~Brun$^{11}$,
P.~Bun\v{c}i\'{c}$^{10,11}$, V.~Cerny$^{4}$, O.~Chvala$^{15}$,
J.G.~Cramer$^{16}$, P.~Csat\'{o}$^{5}$, P.~Dinkelaker$^{10}$,
V.~Eckardt$^{14}$, P.~Filip$^{14}$, H.G.~Fischer$^{11}$,
Z.~Fodor$^{5}$, P.~Foka$^{8}$, P.~Freund$^{14}$,
V.~Friese$^{13}$, J.~G\'{a}l$^{5}$,
M.~Ga\'zdzicki$^{10}$, G.~Georgopoulos$^{2}$, E.~G{\l}adysz$^{7}$, 
S.~Hegyi$^{5}$, C.~H\"{o}hne$^{13}$, 
P.G.~Jones$^{3}$, K.~Kadija$^{11,19}$, A.~Karev$^{14}$,
V.I.~Kolesnikov$^{9}$, T.~Kollegger$^{10}$, M.~Kowalski$^{7}$, 
I.~Kraus$^{8}$, M.~Kreps$^{4}$, M.~van~Leeuwen$^{1}$, 
P.~L\'{e}vai$^{5}$, A.I.~Malakhov$^{9}$,
C.~Markert$^{8}$, B.W.~Mayes$^{12}$, G.L.~Melkumov$^{9}$,
A.~Mischke$^{8}$,J.~Moln\'{a}r$^{5}$, J.M.~Nelson$^{3}$,
G.~P\'{a}lla$^{5}$, A.D.~Panagiotou$^{2}$,
K.~Perl$^{18}$, A.~Petridis$^{2}$, M.~Pikna$^{4}$, L.~Pinsky$^{12}$,
F.~P\"{u}hlhofer$^{13}$,
J.G.~Reid$^{16}$, R.~Renfordt$^{10}$, W.~Retyk$^{18}$,
C.~Roland$^{6}$, G.~Roland$^{6}$, A.~Rybicki$^{7}$,
A.~Sandoval$^{8}$, H.~Sann$^{8}$, N.~Schmitz$^{14}$, P.~Seyboth$^{14}$,
F.~Sikl\'{e}r$^{5}$, B.~Sitar$^{4}$, E.~Skrzypczak$^{18}$,
G.T.A.~Squier$^{3}$, R.~Stock$^{10}$, H.~Str\"{o}bele$^{10}$, T.~Susa$^{19}$,
I.~Szentp\'{e}tery$^{5}$, J.~Sziklai$^{5}$,
T.A.~Trainor$^{16}$, D.~Varga$^{5}$, M.~Vassiliou$^{2}$,
G.I.~Veres$^{5}$, G.~Vesztergombi$^{5}$,
D.~Vrani\'{c}$^{8}$, S.~Wenig$^{11}$, A.~Wetzler$^{10}$, 
I.K.~Yoo$^{13}$, J.~Zaranek$^{10}$, J.~Zim\'{a}nyi$^{5}$
}

\vspace{0.3cm}
\noindent{
$^{1}$NIKHEF, Amsterdam, Netherlands. \\
$^{2}$Department of Physics, University of Athens, Athens, Greece.\\
$^{3}$Birmingham University, Birmingham, England.\\
$^{4}$Comenius University, Bratislava, Slovakia.\\
$^{5}$KFKI Research Institute for Particle and Nuclear Physics, Budapest, Hungary.\\
$^{6}$MIT, Cambridge, USA.\\
$^{7}$Institute of Nuclear Physics, Cracow, Poland.\\
$^{8}$Gesellschaft f\"{u}r Schwerionenforschung (GSI), Darmstadt, Germany.\\
$^{9}$Joint Institute for Nuclear Research, Dubna, Russia.\\
$^{10}$Fachbereich Physik der Universit\"{a}t, Frankfurt, Germany.\\
$^{11}$CERN, Geneva, Switzerland.\\
$^{12}$University of Houston, Houston, TX, USA.\\
$^{13}$Fachbereich Physik der Universit\"{a}t, Marburg, Germany.\\
$^{14}$Max-Planck-Institut f\"{u}r Physik, Munich, Germany.\\
$^{15}$Institute of Particle and Nuclear Physics, Charles University, Prague, Czech Republic.\\
$^{16}$Nuclear Physics Laboratory, University of Washington, Seattle, WA, USA.\\
$^{17}$Institute for Nuclear Studies, Warsaw, Poland.\\
$^{18}$Institute for Experimental Physics, University of Warsaw, Warsaw, Poland.\\
$^{19}$Rudjer Boskovic Institute, Zagreb, Croatia.\\
}


\vspace{0.5cm}

\noindent{
{\bf Abstract:}
Experiment NA49 at the Cern SPS uses a large acceptance detector
for a systematic study of particle yields and correlations in
nucleus--nucleus, nucleon--nucleus and nucleon-nucleon collisions.
Preliminary results for Pb+Pb collisions at 40, 80 and 158 A$\cdot$GeV 
beam energy are shown and compared to measurements at lower 
and higher energies.
} 

\vspace{0.5cm}
\section{Introduction}
The primary purpose of the heavy ion programme at the CERN SPS is 
the search for evidence of a transient deconfined state
of strongly interacting matter during the early stage of 
nucleus--nucleus collisions \cite{qm99}. The transition from
a dilute state of individual hadrons to a phase of quasi-free
quarks and gluons, the quark gluon plasma (QGP), was first
suggested using qualitative arguments \cite{qgp} and later
confirmed by quantum chromodynamics (QCD) on the lattice provided the
energy density reaches sufficiently high values in an extended
volume.

\begin{figure}[hbt]
\epsfig{file=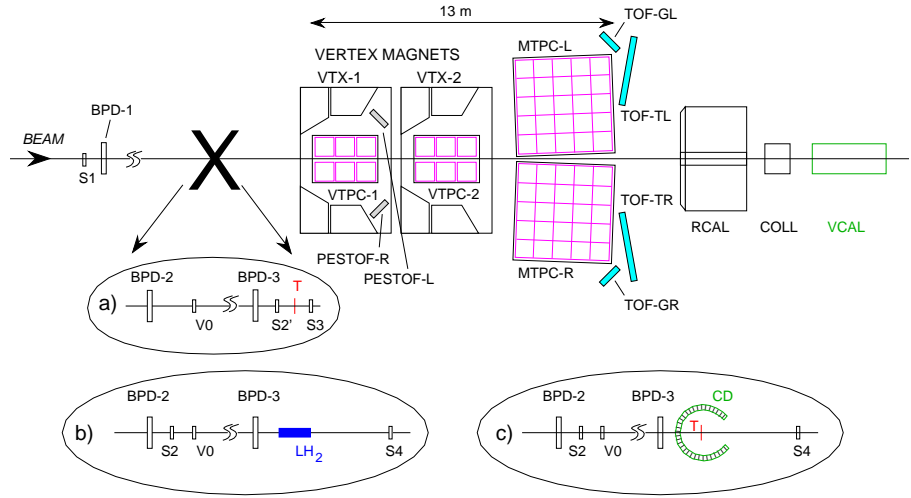,width=12.0cm}
\caption{Schematic layout of the NA49 experiment at the Cern SPS showing
beam detectors, superconducting dipole magnets, time projection
chambers (VTPC,MTPC), time-of-flight arrays (TOF) and calorimeters (RCAL,VCAL).
A thin solid target T is used for A+A collisions (a), which is surrounded by
a detector of slow protons (CD) for p+A collsions (c). A liquid H$_2$ target is
employed for p+p collisions (b)}
\label{fig_na49}
\end{figure}

The results from the heavy-ion program at CERN in fact indicated 
that deconfinement may set in within the SPS energy 
range \cite{qgpcern}. Within most model scenarios the data imply that
the initial energy density exceeds the critical value.
Originally proposed signatures \cite{qgpsign} of the QGP were observed 
in Pb+Pb collisions at the top SPS energy, i.e. $J/\Psi$ suppression,
strangeness enhancement, and possibly thermal photons and dileptons.
The significance of these signals as QGP signatures has come under renewed 
scrutiny. Moreover, there is no observational evidence for a sharp phase
transition from QGP to hadrons such as phase coexistence \cite{hu95},
critical fluctuations \cite{hu95,step99} or more speculative effects like
DCC \cite{raj93} or parity violation \cite{khar99}.

The NA49 experiment is performing an energy scan from 20--158 A$\cdot$GeV
in an effort to strengthen the evidence for the onset of deconfinement 
by searching for anomalies in the energy dependence of experimental
observables. Data at 40, 80 and 158 A$\cdot$GeV have so far been recorded and
analysed. A brief description of the apparatus (section 2) is followed 
by preliminary results for $\pi$, K, $\Lambda$ and $\bar{\Lambda}$ yields,
$\pi\pi$ correlations, event-to-event charge fluctuations, anisotropic flow
(sections 3 - 6), and conclusions (section 7).

\section{The NA49 detector}\label{detector}

The NA49 experiment \cite{na49_nim} was designed for the investigation
of hadron production in the most violent Pb+Pb interactions at the Cern
SPS. The main features (Fig.\ref{fig_na49}) are large acceptance precision tracking 
and particle identification using time projection chambers (TPCs). The first two are
located inside superconducting dipole magnets which provide the particle
trajectory bending necessary for momentum determination. Charged particles
in the forward hemisphere of the reaction are identified from the measurement 
of their energy loss dE/dx in the TPC gas (accuracy 3 --4 \%). 
At central rapidity the identification is further
improved by measurement of the time-of-flight (resolution 60 ps) to arrays of
scintillation counter tiles (TOF-T) and strips (TOF-G). Strange particles
are detected via decay topology and invariant mass measurement. The
forward calorimeter VCAL measures the energy of the projectile spectators from
which one can deduce the impact parameter in A+A collisions. 

\begin{figure}[hbt]
\begin{center}
\mbox{
 \parbox{8.0cm}{
  \epsfig{figure=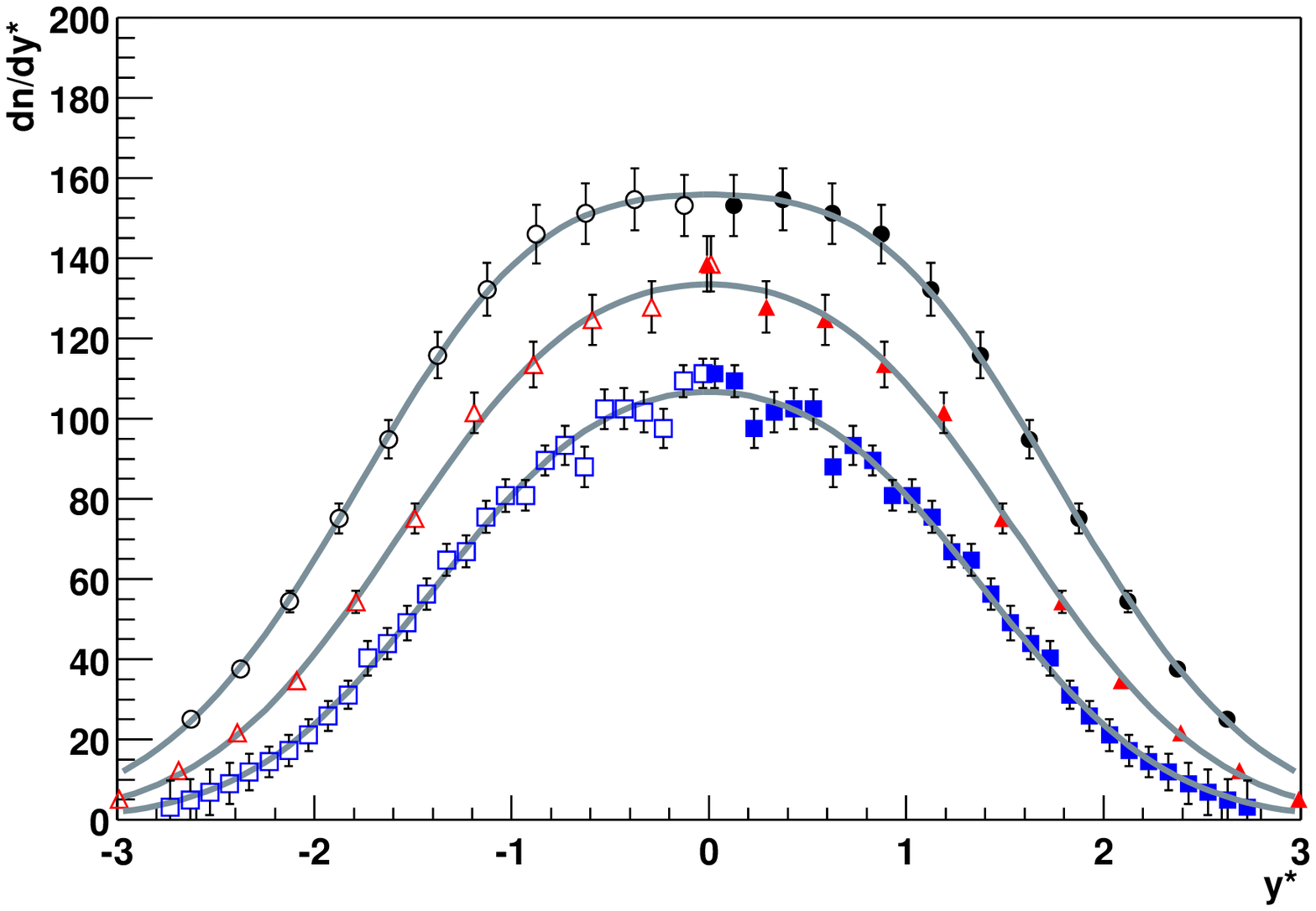,width=8.0cm}}
 \parbox{7.0cm}{
  \epsfig{figure=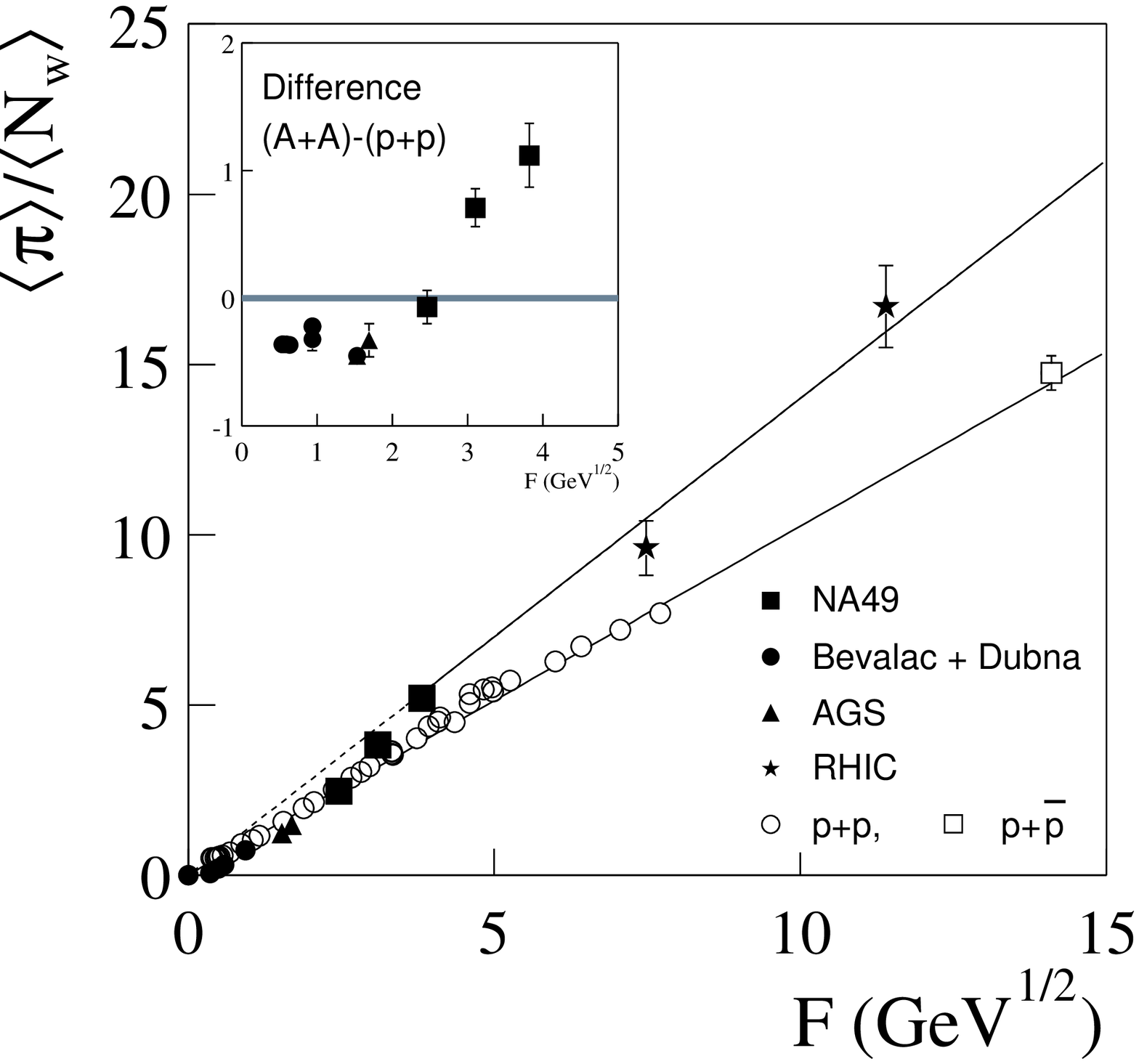,width=7.0cm}}
}
\end{center}
\caption{
Left: rapidity distribution of $\pi^-$ in central Pb+Pb collisons at 40 (squares),
80 (triangles) and 158 (dots) A$\cdot$GeV. Open symbols show values reflected at 
y$^{\star}$=0 (NA49 preliminary). Right: total pion multiplicity $\langle \pi \rangle$ 
produced per wounded (participant) nucleon versus the Fermi energy variable 
F $\approx s_{NN}^{0.25}$ for p+p reactions (open symbols) and central
nucleus--nucleus collisions (full symbols).
}
\label{fig_pions}
\end{figure}

\section{Yields of $\pi$, K, $\Lambda$ and $\bar{\Lambda}$}\label{yields}

Raw K$^+$ and K$^-$ yields were extracted from fits of the distributions of dE/dx
and TOF (where available) in narrow bins of rapidity y and transverse momentum 
p$_T$. The resulting spectra were then corrected for geometrical acceptance,
losses due to in--flight decays and reconstruction efficiency. The latter was
determined from embedding simulated tracks into real events and amounted to 
$\approx$ 95\%. Spectra of $\pi^-$ mesons were derived from the acceptance
corrected negatively charged particle yields in p$_T$ and y (assuming the $\pi$
mass) by subtracting the estimated contribution of K$^-$, $\bar{\textrm{p}}$ and 
the contamination from secondary hadron decays. The ratio $\pi^+$/$\pi^-$ was
determined in the region where both dE/dx and TOF are available (0.91, 0.94 and 0.97
at 40, 80 and 158 A$\cdot$GeV) and was assumed to be y independent.

\begin{figure}[hbt]
\begin{center}
\mbox{
 \parbox{5.0cm}{
  \epsfig{figure=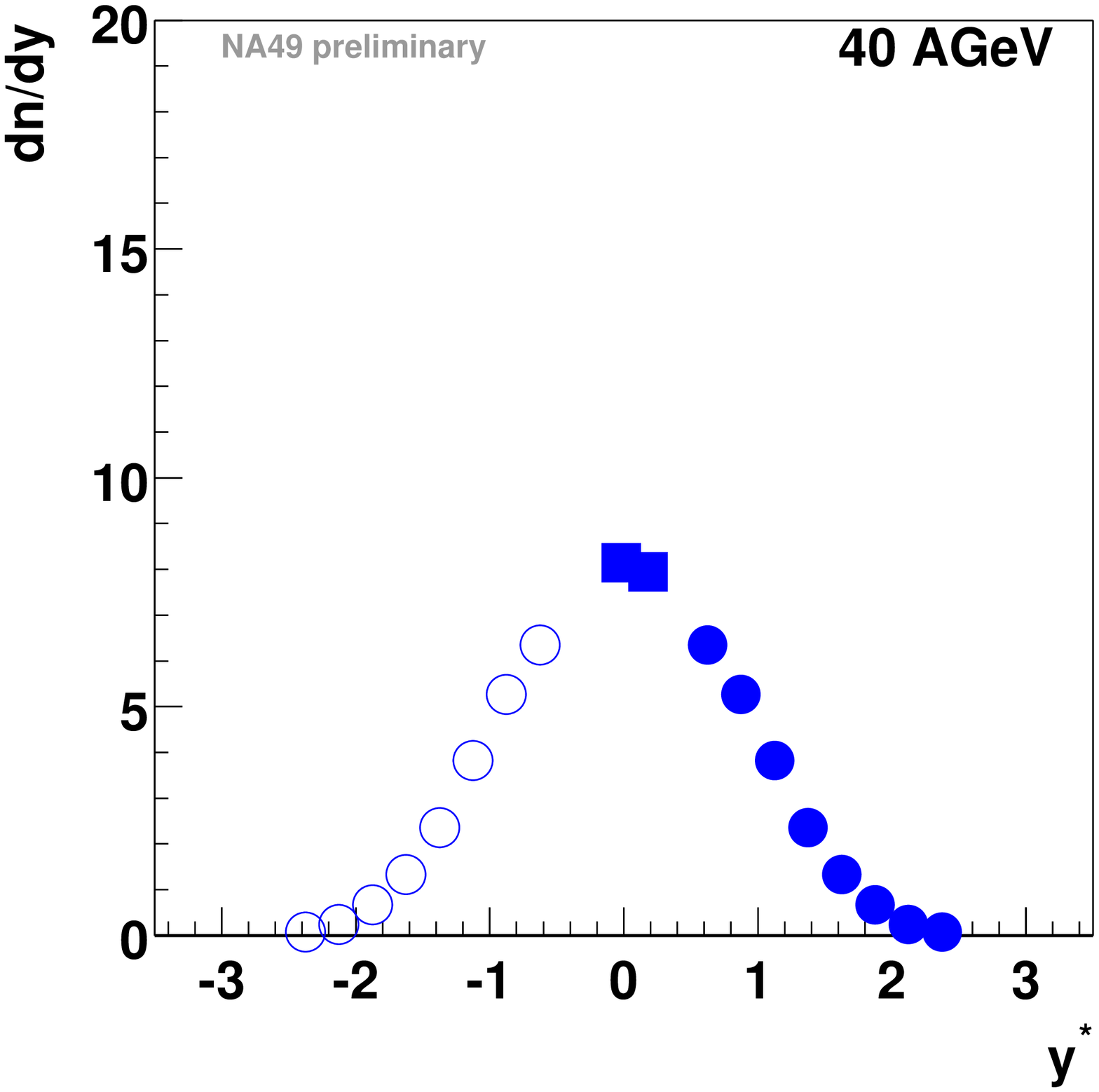,width=5.0cm}}
 \parbox{5.0cm}{
  \epsfig{figure=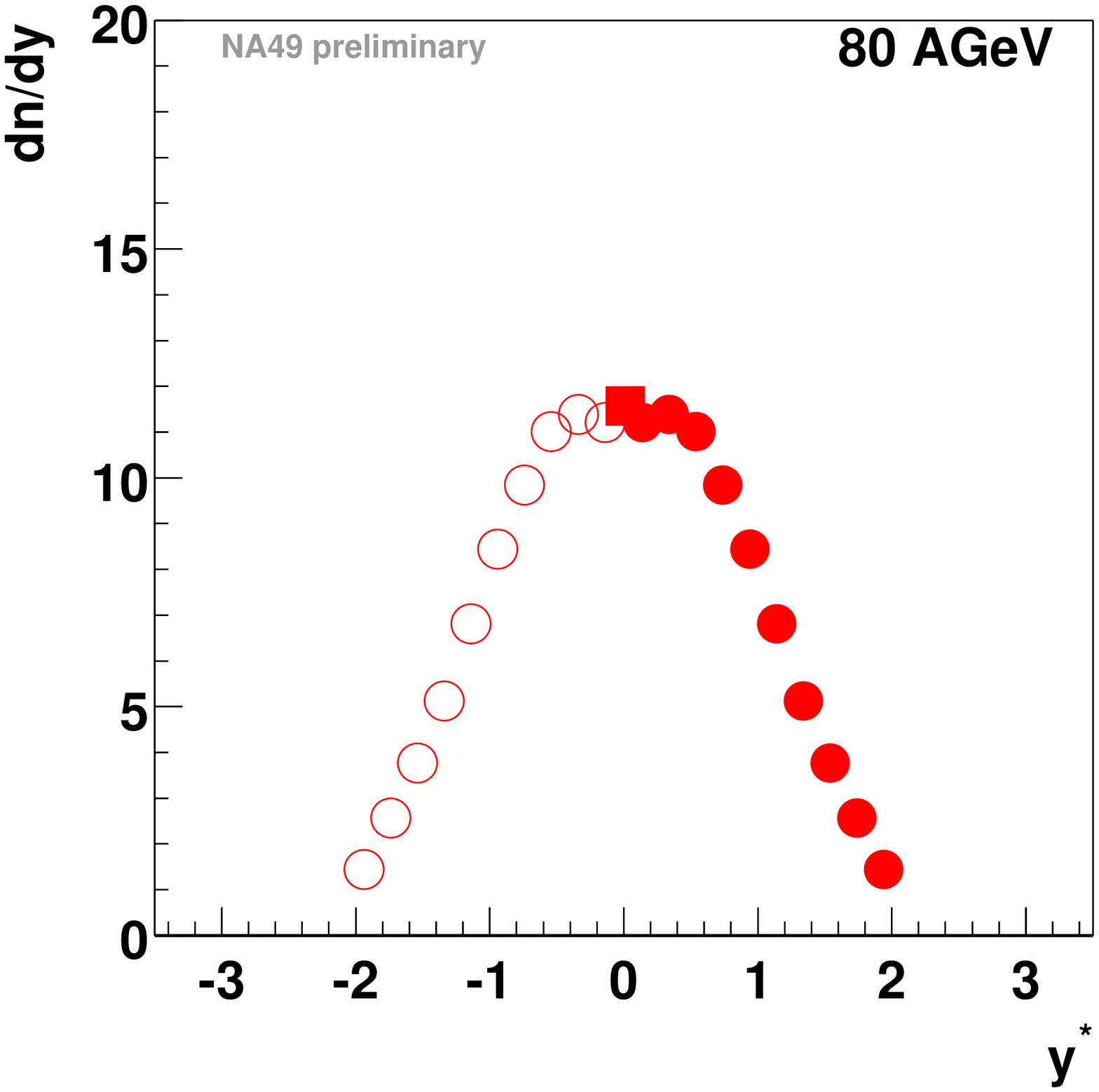,width=5.0cm}}
 \parbox{5.0cm}{
  \epsfig{figure=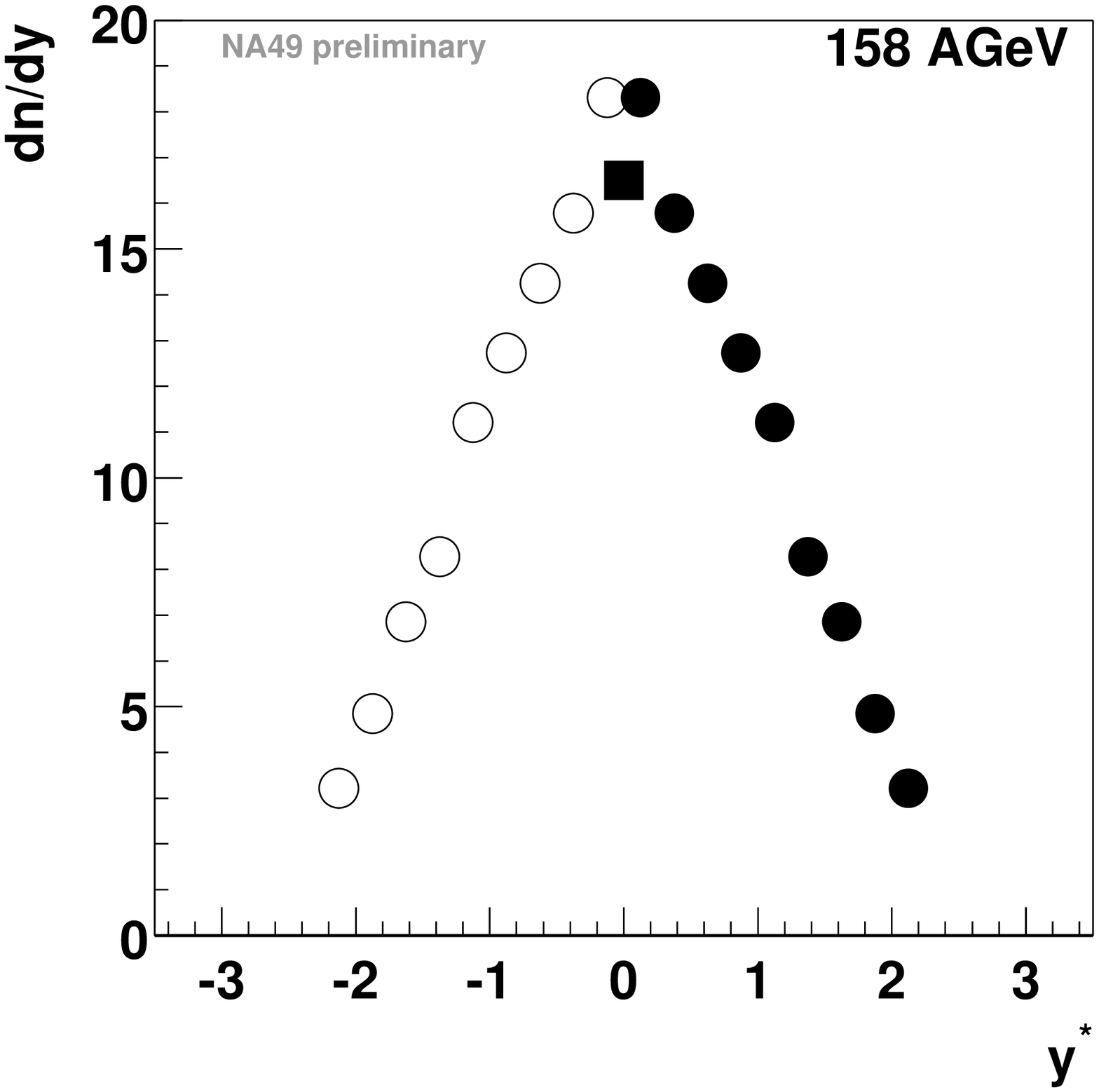,width=5.0cm}}
}
\mbox{
 \parbox{5.0cm}{
  \epsfig{figure=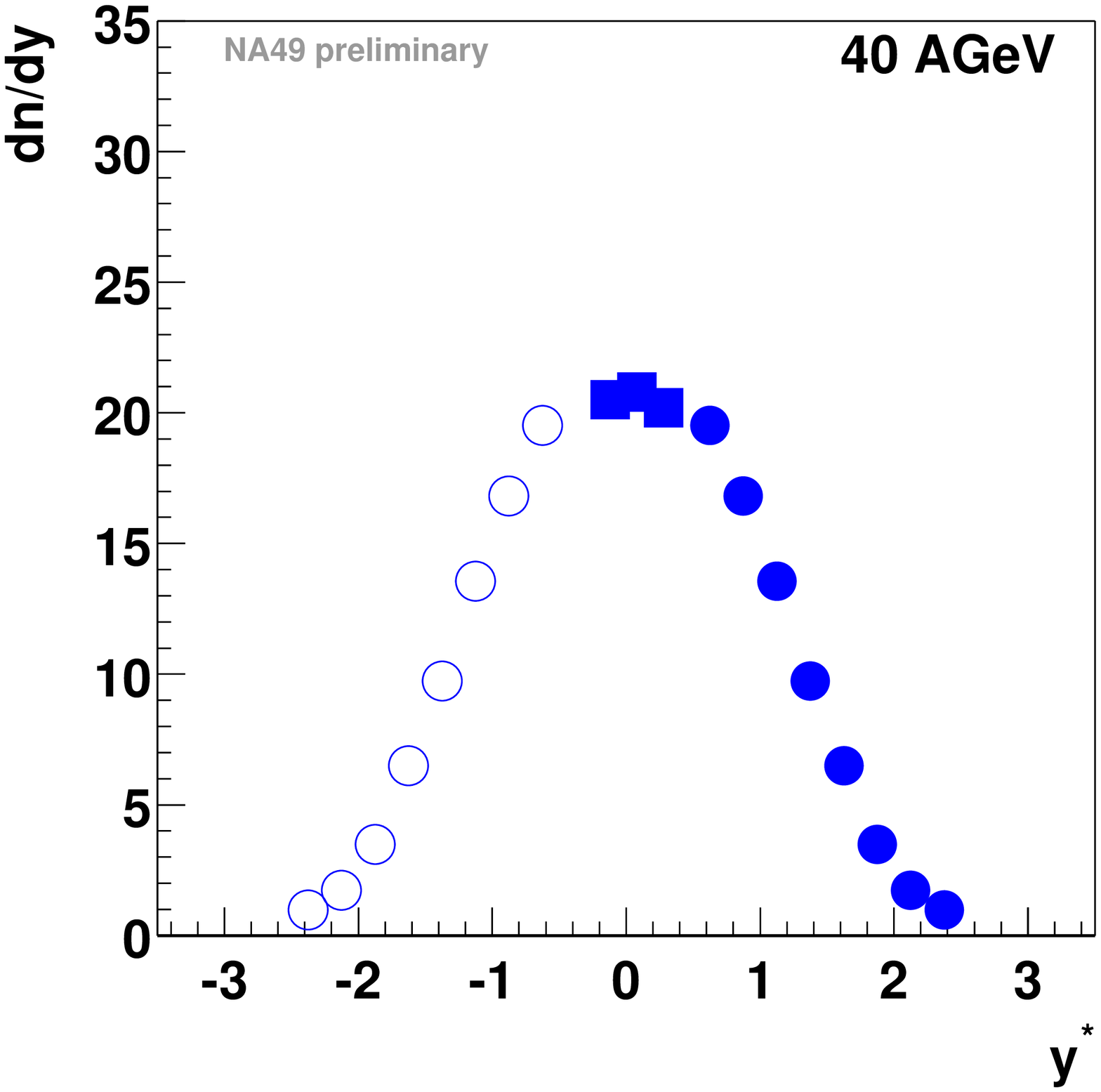,width=5.0cm}}
 \parbox{5.0cm}{
  \epsfig{figure=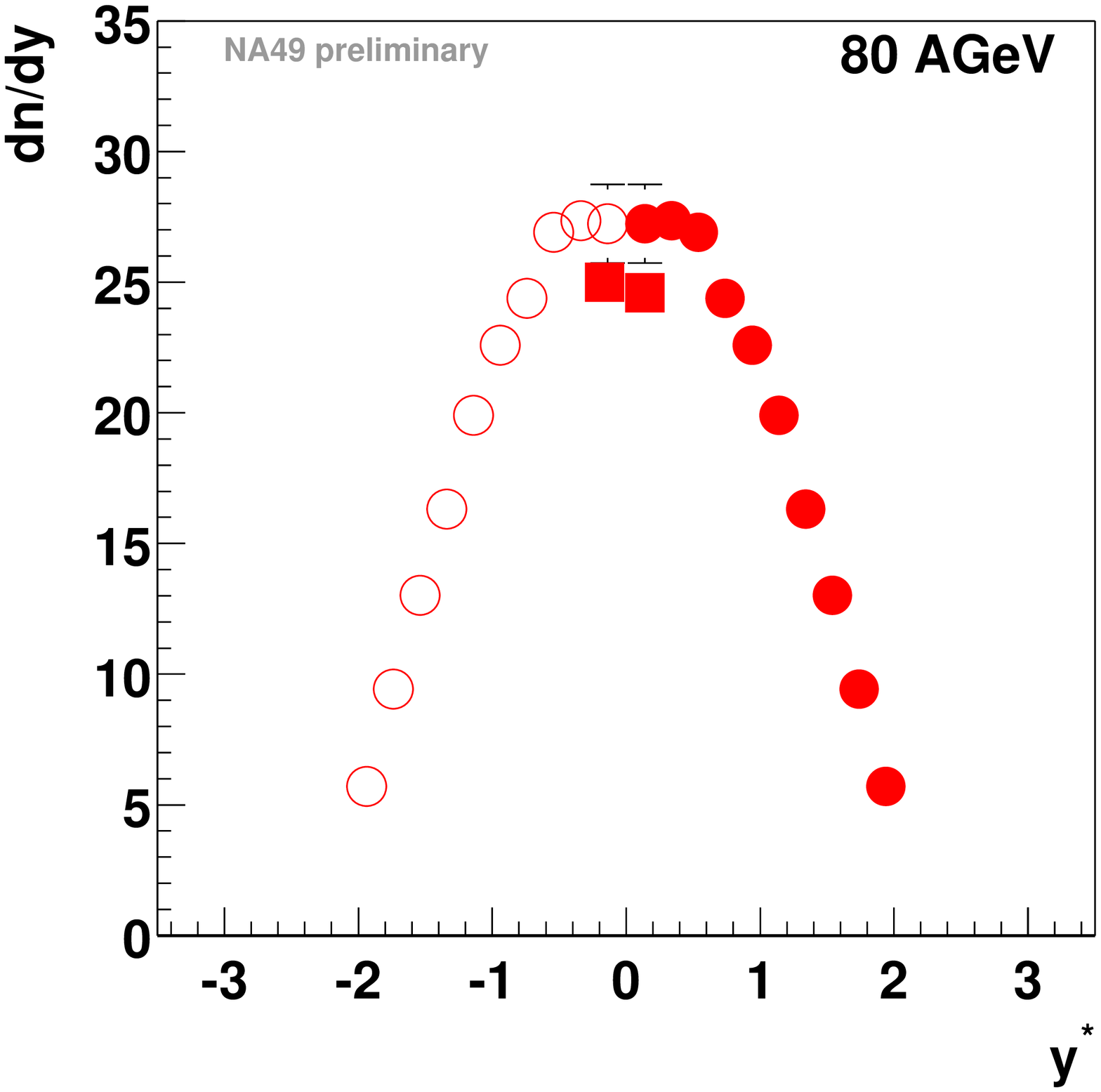,width=5.0cm}}
 \parbox{5.0cm}{
  \epsfig{figure=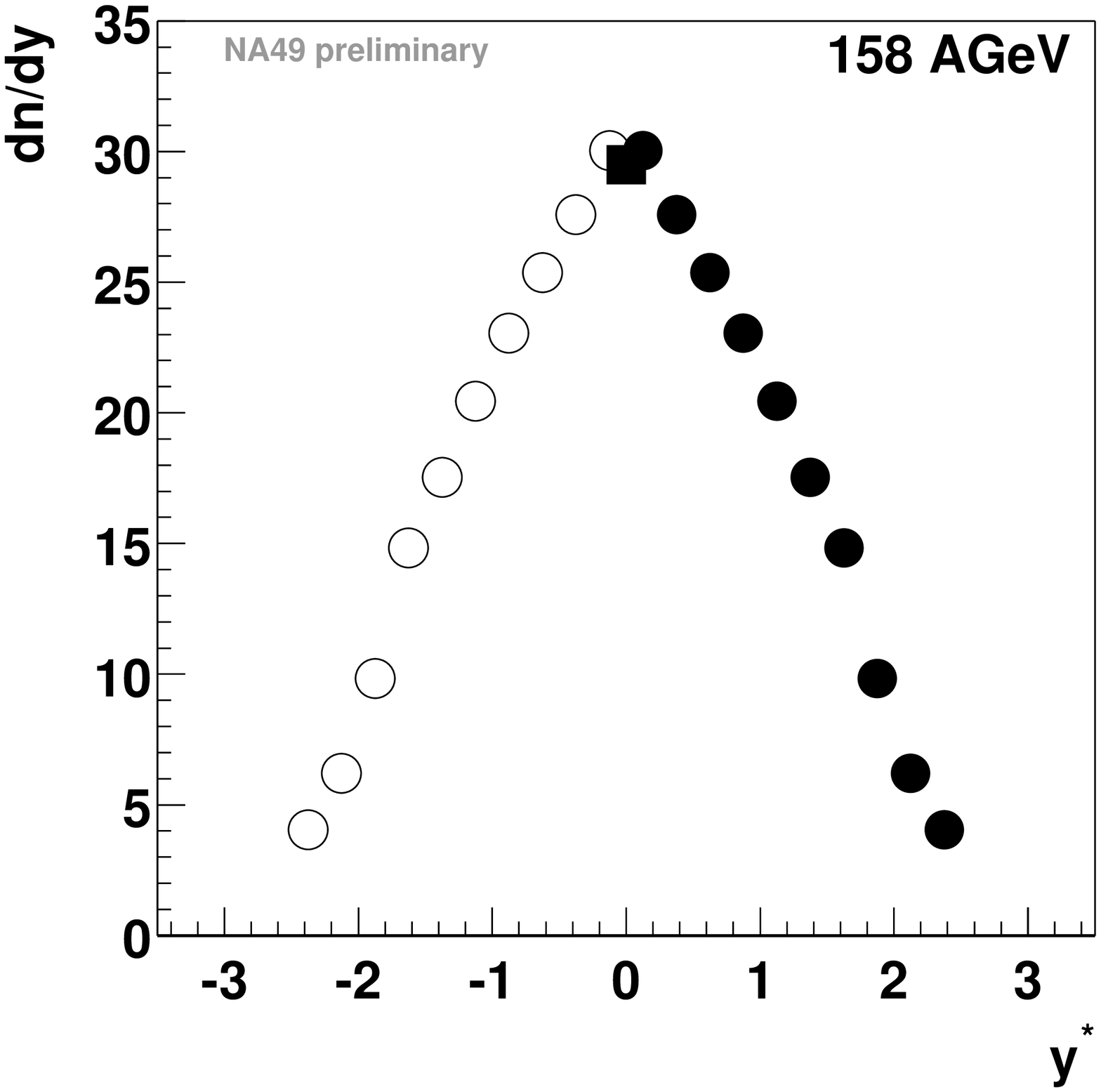,width=5.0cm}}
}
\end{center}
\caption{
Rapidity distribution of K$^-$ (top row) and K$^+$ (bottom row) from
dE/dx (circles) and combined dE/dx and TOF (squares) analysis. Open
symbols show values reflected at y$^{\star}$=0. (NA49 preliminary)
}
\label{fig_kaons}
\end{figure}

The resulting rapidity distributions of $\pi^-$ are plotted in Fig.\ref{fig_pions}, left.
The integrated yields are 312$\pm$15, 445$\pm$22 and 610$\pm$30 at 40, 80 and
158 A$\cdot$GeV respectively. Pions are the dominant produced particle species and
thus their number provides a measure of the entropy in a statistical model description of the
reaction. The yield of pions (estimated here as 
$\langle \pi \rangle = 1.5 \cdot (\langle \pi^- \rangle + \langle \pi^+ \rangle )$)
divided by the number of wounded nucleons (participants) N$_W$ is shown versus the Fermi
energy variable F $\equiv (\sqrt{s} - 2 m_N)^{3/4}/\sqrt{s_{NN}}^{1/4} \approx s_{NN}^{1/4}$
in Fig.\ref{fig_pions}, right. While p+p data show a linear rise throughout
there is a change for A+A collisions (illustrated more clearly in the inset) in the SPS
energy range. Below one finds a regime of slight suppression, above a region of 
enhancement with a steeper linear rise than for p+p reactions. This steepening has
been interpreted as indicating the activation of a large number of partonic degrees
of freedom at the onset of deconfinement \cite{GaGo99}.

\begin{figure}[hbt]
\begin{center}
\mbox{
 \parbox{7.0cm}{
  \epsfig{figure=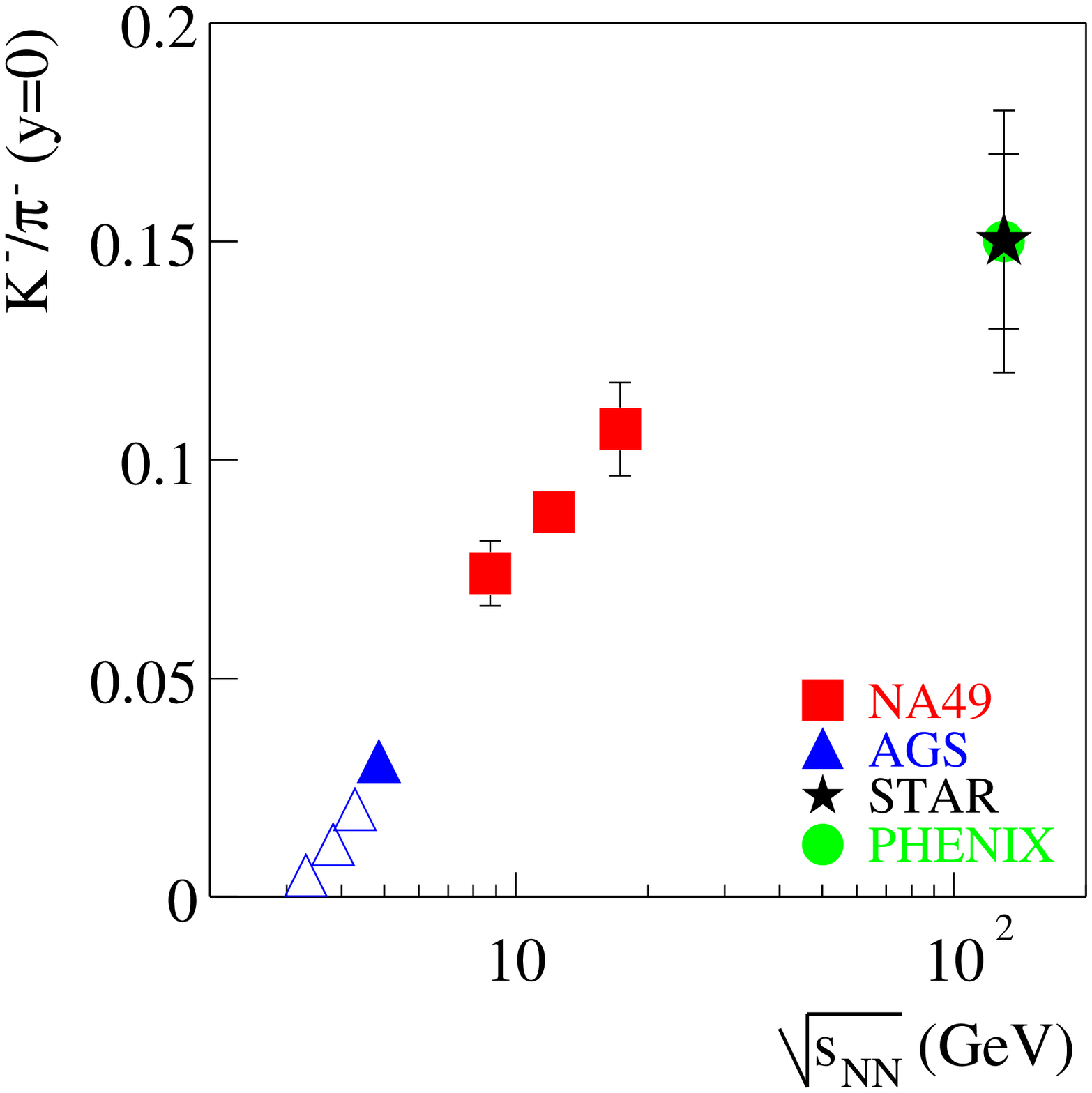,width=7.0cm}}
 \parbox{7.0cm}{
  \epsfig{figure=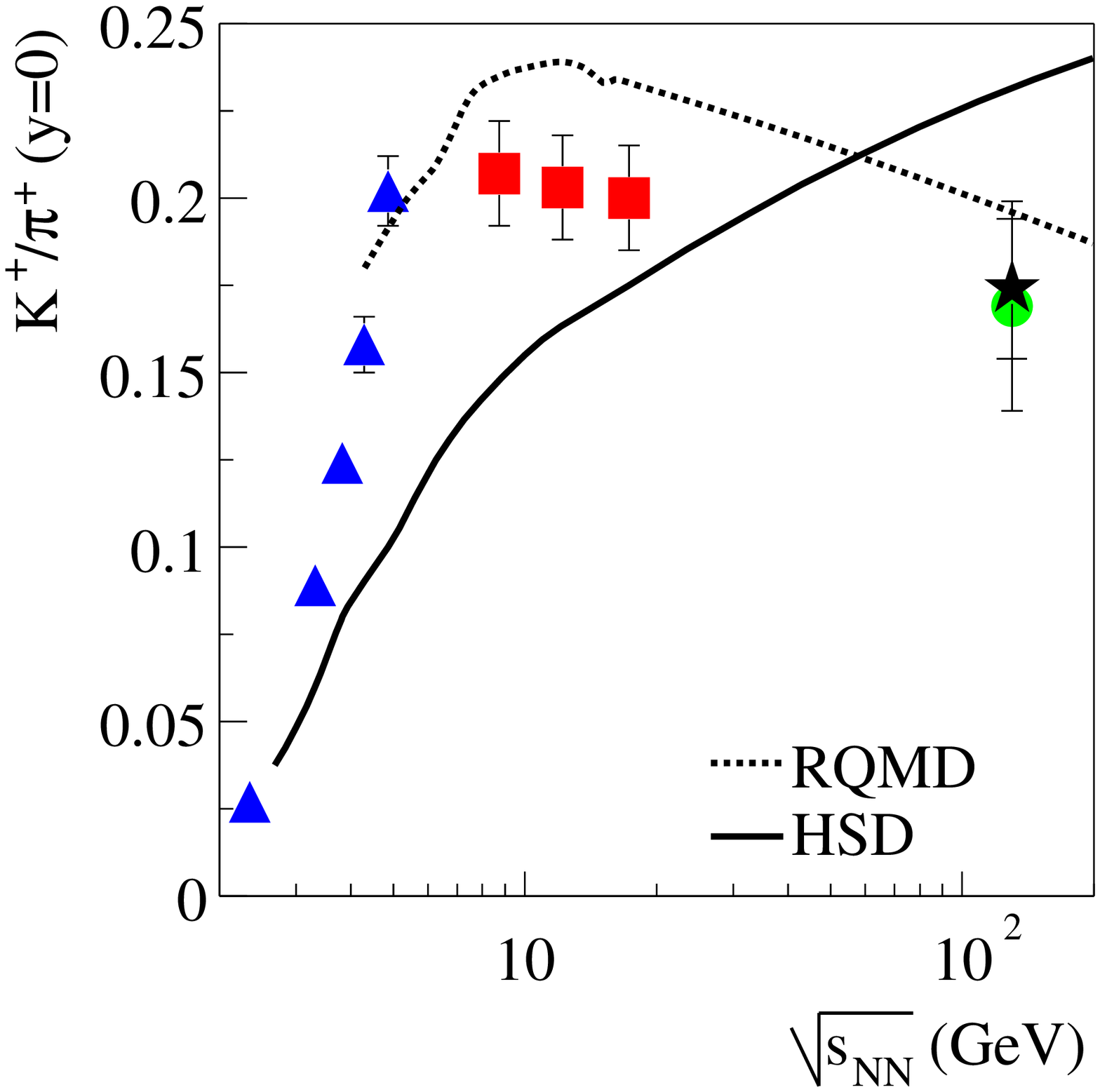,width=7.0cm}}
}
\end{center}
\caption{
Midrapidity ratio of K$^-$/$\pi^-$ (left) and K$^+$/$\pi^+$ (right) as function of
energy from NA49 (squares, preliminary) compared to measurements at lower and
higher energies. Predictions of the RQMD \cite{RQMD00} (dotted) and HSD \cite{HSD00}
(full curve) models are shown.
}
\label{fig_kaons_r}
\end{figure}

Kaons contain about 75\% of the $s,\bar{s}$ quarks in the produced hadrons at SPS energies
and thus their number indicates the total strangeness content of the final state. The
rapidity distributions are displayed in Fig.\ref{fig_kaons} and integrate with small
extrapolation to total yields of 18$\pm$1, 29$\pm$2, 50$\pm$5 for K$^-$ and 
56$\pm$3, 79$\pm$5, 95$\pm$9 for K$^+$ at 40, 80, 160 A$\cdot$GeV respectively. Yields
of most particles, of course, increase with energy. Changes in the composition of
the produced system are better characterised by particle ratios. Measurements of
K$^-$/$\pi^-$ and K$^+$/$\pi^+$ at midrapidity from NA49 are plotted versus energy 
in Fig.\ref{fig_kaons_r} and compared to results at lower and higher energies.

\begin{figure}[hbt]
\begin{center}
\mbox{
 \parbox{8.0cm}{\epsfig{figure=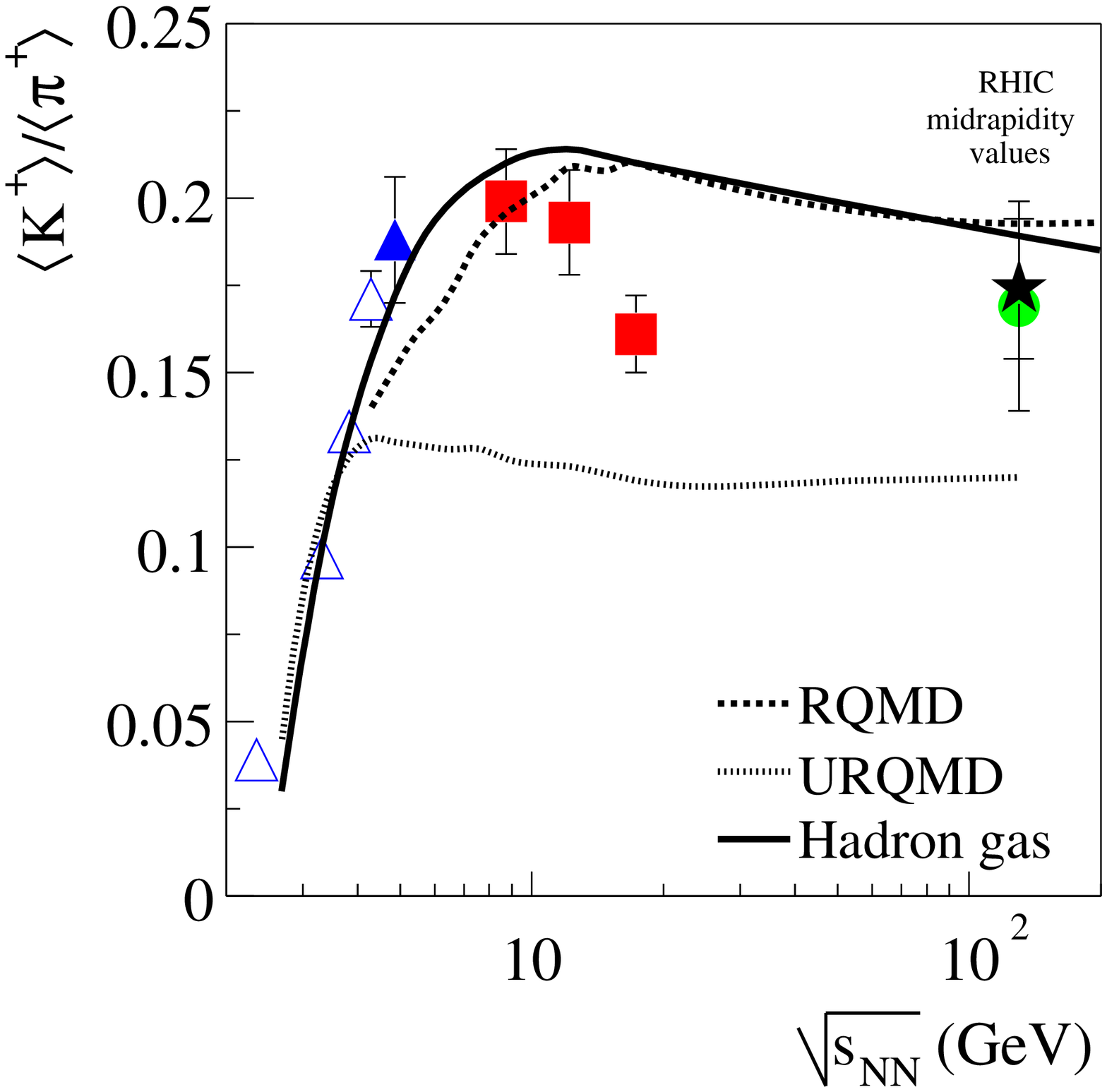,width=8.0cm}}
 \parbox{8.0cm}{\vspace*{-0.8cm}\epsfig{figure=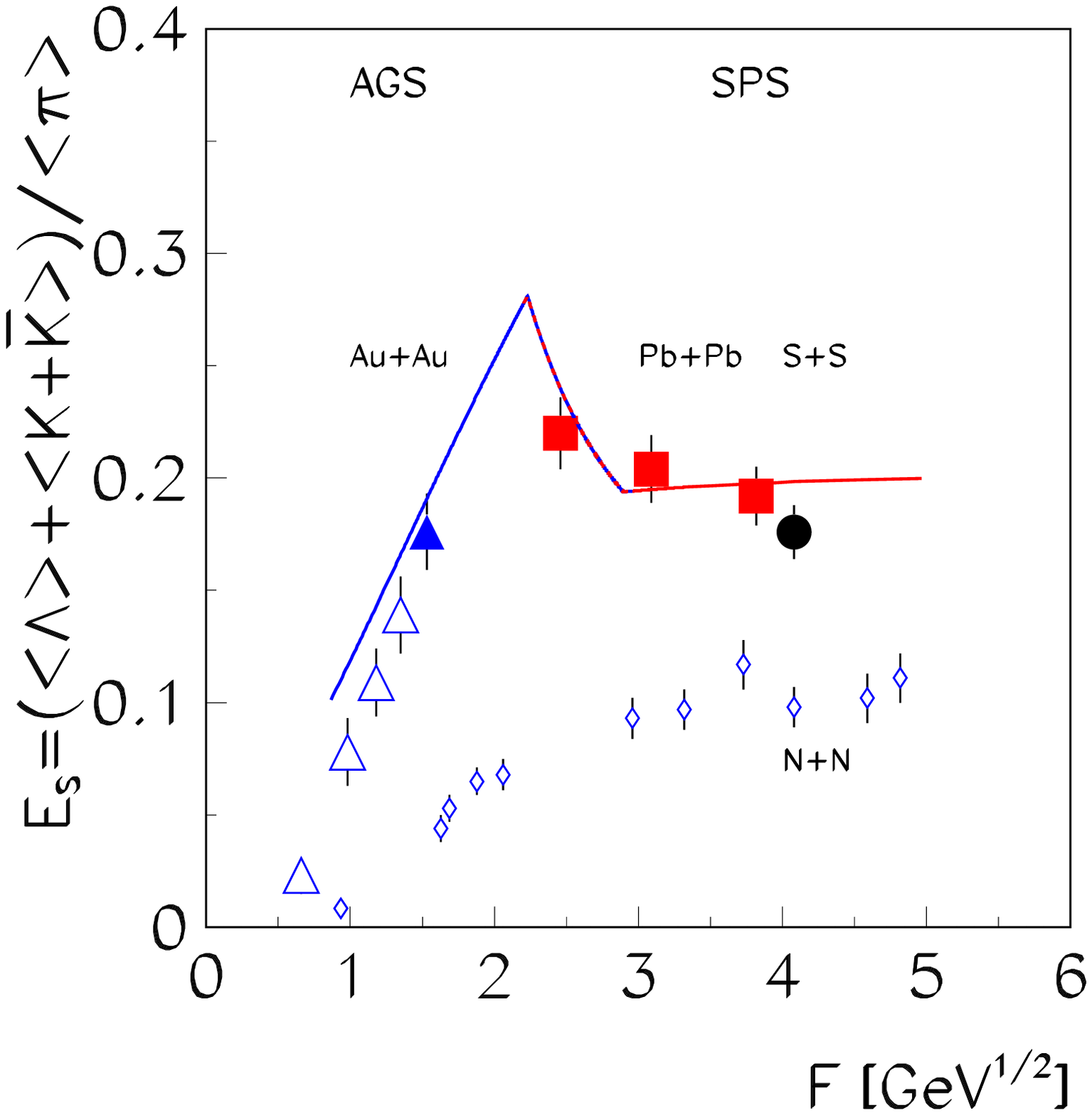,width=8.0cm}}
}
\end{center}
\caption{
Left: 4$\pi$ ratio $\langle$K$^+\rangle$/$\langle\pi^+\rangle$ versus energy compared
to predictions of the RQMD \cite{RQMD00} (dashed), UrQMD \cite{URQMD98} (dotted) and extended 
statistical \cite{ClRe01} (solid curve) models. Right: the strangeness content measure 
E$_s$ versus the Fermi energy variable F$\approx s_{NN}^{0.25}$ compared to the 
prediction of the statistical model of the early stage \cite{GaGo99} (curves).
}
\label{fig_model}
\end{figure}

A continuous increase is seen for K$^-$/$\pi^-$. For K$^+$/$\pi^+$ one finds a steeper rise 
followed by a maximum at the lower end of the SPS energy region and a gradual decrease.
Within a reaction scenario based on nucleon--nucleon collisions these features might 
be attributed to thresholds and the decrease of the baryon density with increasing energy. 
The ratio K$^-$/$\pi^-$ exhibits the threshold of the K$\bar{\textrm{K}}$ production mechanism. 
The lower mass threshold of associate KY production leads to a steeper rise in
K$^+$/$\pi^+$ and the rapidly falling baryon density may result in a compensation of the 
declining contribution from the KY by the growing contribution from the K$\bar{\textrm{K}}$ 
production mechanism. However, the continuous increase seen for nucleon--nucleon collisions
(e.g. Fig.\ref{fig_model}, right) does not really support such an interpretation.

\begin{figure}[hbt]
\begin{center}
\mbox{
 \hspace*{-0.5cm}\parbox{8.0cm}{\epsfig{figure=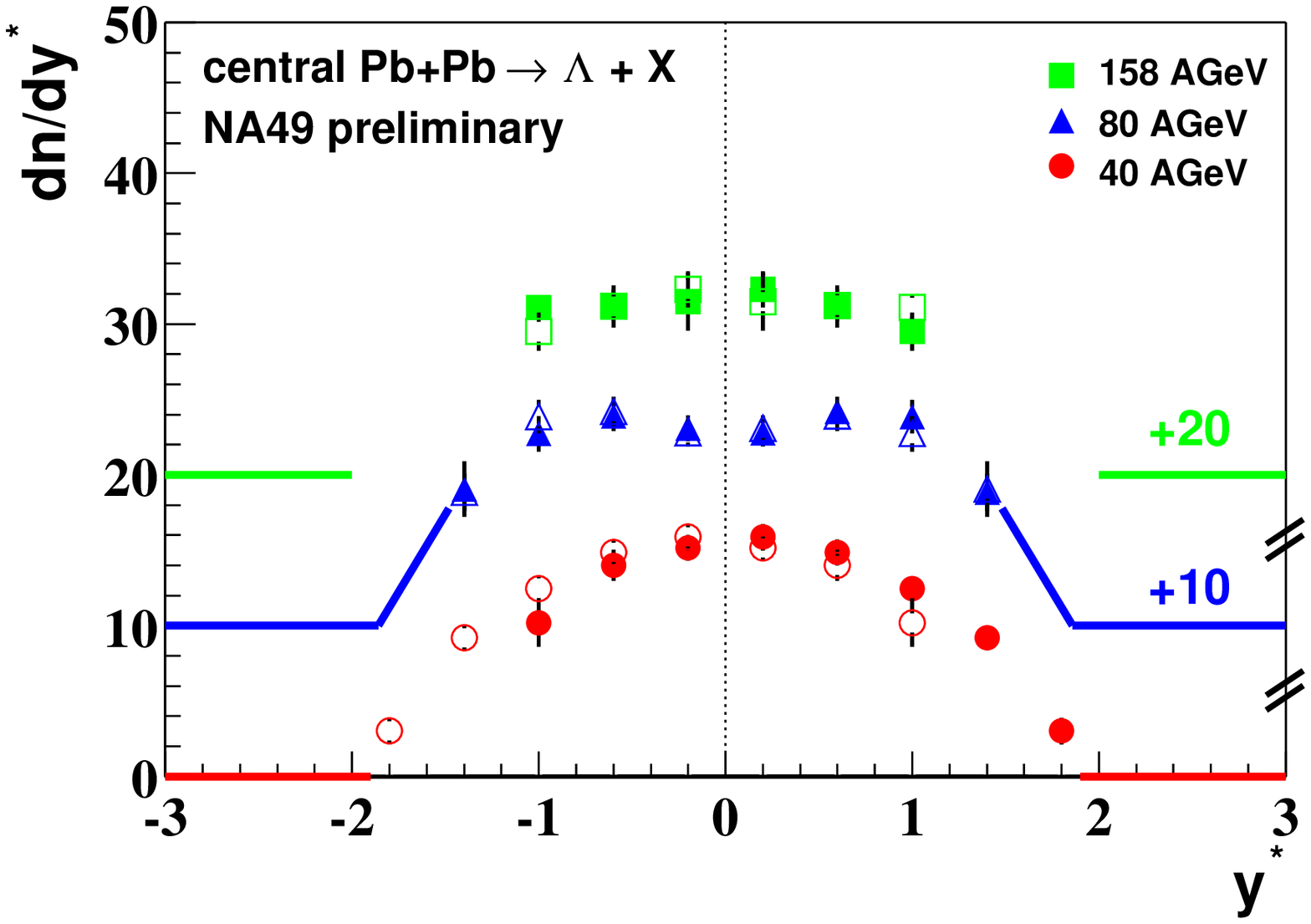,width=8.0cm}}
 \hspace*{-0.5cm}\parbox{8.5cm}{\epsfig{figure=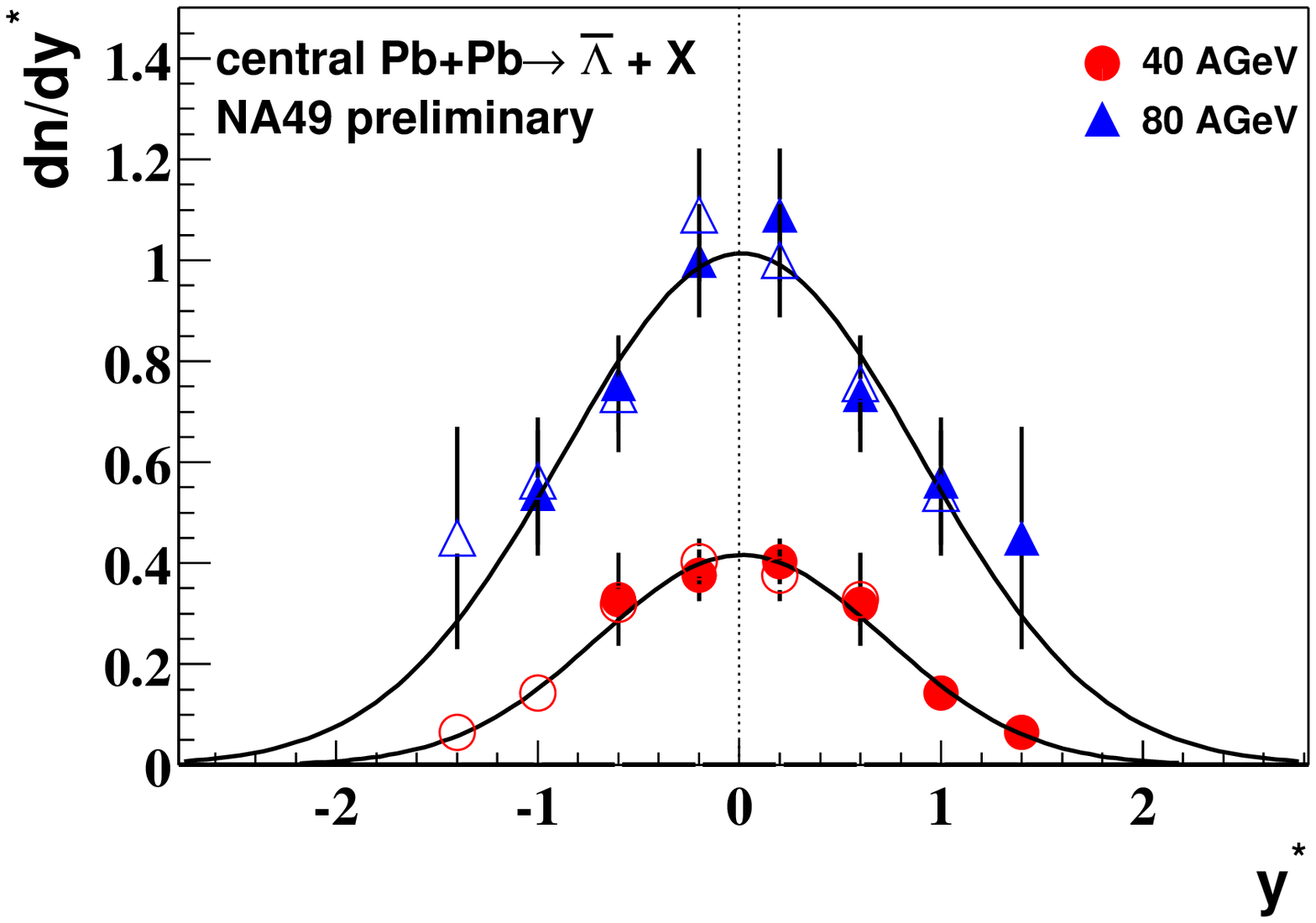,width=8.5cm}}
}
\end{center}
\caption{
Rapidity distribution of $\Lambda$ (left) and $\bar{\Lambda}$ (right). Open
symbols show values reflected at y$^{\star}$=0. $\Lambda$ yields at 80 and
158 A$\cdot$GeV are displaced vertically by 10 respectively 20 units for clarity.
(NA49 preliminary)
}
\label{fig_llbar}
\end{figure}

\begin{figure}[hbt]
\begin{center}
\mbox{
 \hspace*{-0.5cm}\parbox{8.5cm}{\epsfig{figure=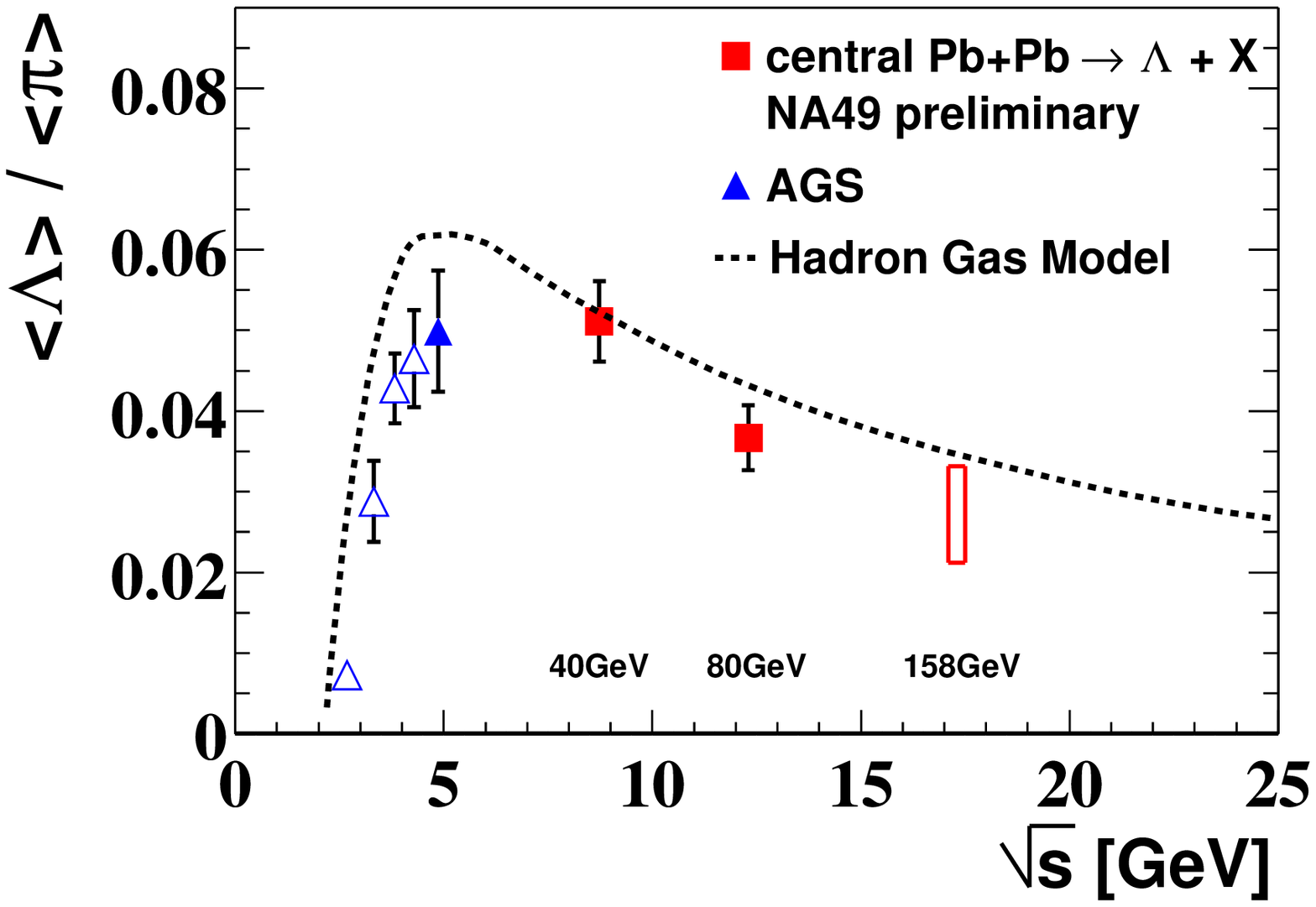,width=8.5cm}}
 \hspace*{-0.5cm}\parbox{8.5cm}{\epsfig{figure=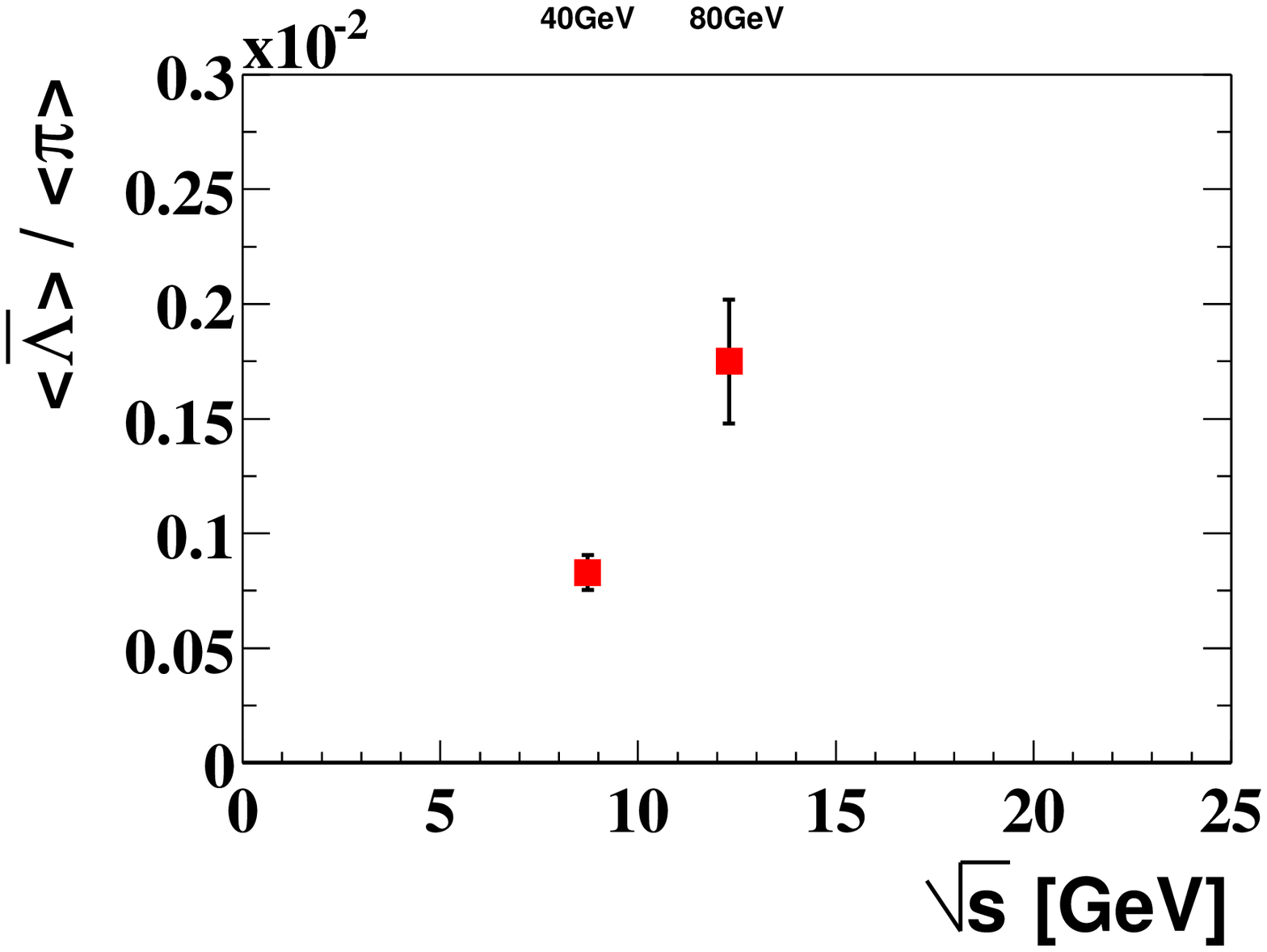,width=8.5cm}}
}
\end{center}
\caption{
4$\pi$ yield ratios $\langle\Lambda\rangle$/$\langle\pi\rangle$ (left) and
$\langle\bar{\Lambda}\rangle$/$\langle\pi\rangle$ (right) versus energy 
from NA49 (squares, preliminary) and lower energy AGS experiments. The
dotted line shows a prediction from the extended hadron gas model \cite{ClRe01}.
}
\label{fig_llbar_r}
\end{figure}

The most abundantly produced hyperons are $\Lambda$ and $\bar{\Lambda}$ for
which Fig.\ref{fig_llbar} shows the rapidity distributions. These show a broad shape
for $\Lambda$ reflecting the associate production mechanism and the partially stopped
participant nucleon distribution. In contrast the distribution is of narrower 
Gaussian type for $\bar{\Lambda}$ which are most likely produced as members of 
hyperon--antihyperon (Y$\bar{\textrm{Y}}$) pairs. The ratios 
$\langle\Lambda\rangle$/$\langle\pi\rangle$ and
$\langle\bar{\Lambda}\rangle$/$\langle\pi\rangle$ of 4$\pi$ integrated
yields are displayed as a function of energy in Fig.\ref{fig_llbar_r}. One
observes a steep threshold rise for the 
$\langle\Lambda\rangle$/$\langle\pi\rangle$ ratio followed by a decline which 
can be mainly attributed to the rapidly decreasing net baryon density. 
The ratio $\langle\bar{\Lambda}\rangle$/$\langle\pi\rangle$ exhibits a continuous 
rise due to the high mass Y$\bar{\textrm{Y}}$ threshold.
 
Microscopic dynamical and statistical models have been used extensively to describe particle
yields in a wide variety of reactions. The first class is often based on string excitation,
fusion and hadronisation (e.g. HSD \cite{HSD00}, RQMD \cite{RQMD00}, UrQMD \cite{URQMD98}) 
followed by reinteractions of the formed hadrons (RQMD, UrQMD). As seen from 
Fig.\ref{fig_kaons_r} HSD does not reproduce the energy dependence of
the midrapidity K$^+$/$\pi^+$ ratio. On the other hand, RQMD correctly predicts the 
trend, but somewhat overpredicts the ratio in the SPS energy range.

A comparison with the 4$\pi$ ratio $\langle$K$^+\rangle$/$\langle\pi^+\rangle$ is presented 
in Fig.\ref{fig_model} left. Both UrQMD and RQMD get the steep threshold rise. UrQMD 
values are much too low in the plateau due to an overprediction of pions. RQMD does 
not follow the drop in the SPS range which is indicated by the measurements.

Since antihyperon production rates are small and isospin symmetry approximately holds
($\langle$K$^+\rangle \approx \langle$K$^0\rangle$) nearly half of the $\bar{s}$ quarks
in the produced hadrons are contained in K$^+$ mesons. Moreover, strangeness conservation
requires $\langle s \rangle = \langle \bar{s} \rangle$. Thus $\langle$K$^+\rangle$
measures to a good approximation one quarter of all $s$ and $\bar{s}$ quarks in the
final state hadrons. The energy dependence of the $\langle$K$^+\rangle$/$\langle\pi^+\rangle$ 
ratio (see Fig.\ref{fig_model} left) therefore indicates a maximum in the fraction of
strangeness carrying particles in the lower SPS energy range. 

Statistical models have been surprisingly successful in describing ratios of particle
yields in many types of reactions over a wide energy range. Since the widths of rapidity
distributions depend on particle mass in the SPS energy range and below, the model
should preferentially be compared to 4$\pi$ yields. Fits of this model to
NA49 data \cite{becat98} have indicated that while there is relative hadro--chemical equilibrium
in the strange particle sector there seems to be an overall undersaturation with
respect to non-strange particle yields. The statistical model as such makes no prediction
concerning the energy dependence of particle production. However, it has recently been 
supplemented \cite{ClRe01} by a parameterisation of the energy dependence of its two 
main parameters, baryochemical
potential $\mu_B$ and temperature T. Predictions are shown in Fig.\ref{fig_model} left for
$\langle$K$^+\rangle$/$\langle\pi^+\rangle$ and Fig.\ref{fig_llbar_r} right for
$\langle\Lambda\rangle$/$\langle\pi\rangle$. The trend of the data is well reproduced
by this extended statistical model calculation. In detail the decrease in the SPS
energy range of both $\langle$K$^+\rangle$/$\langle\pi^+\rangle$ and
$\langle\Lambda\rangle$/$\langle\pi\rangle$ and even more so of
$\langle\Xi\rangle$/$\langle\pi\rangle$ (not shown) is not well described. 
 
Predictions have also been published for a statistical model which explicitly assumes
a deconfined phase in the early stage above a certain threshold energy \cite{GaGo99}.
In this model the strangeness to entropy ratio is assumed to be established initially 
and to persist through the hadronisation stage. A measure of this quantity is the ratio 
E$_s = ( \langle\Lambda\rangle + \langle K + \bar{K} \rangle )/\langle\pi\rangle$ 
evaluated from the final hadron multiplicities which is plotted in 
Fig.\ref{fig_model} right. After a rise corresponding to the purely hadronic reaction
in the model, one observes a saturation at the a level consistent with the strangeness
to entropy ratio expected for an initially deconfined system. 
 
\begin{figure}[hbt]
\begin{center}
\mbox{
 \parbox{7.0cm}{
  \epsfig{figure=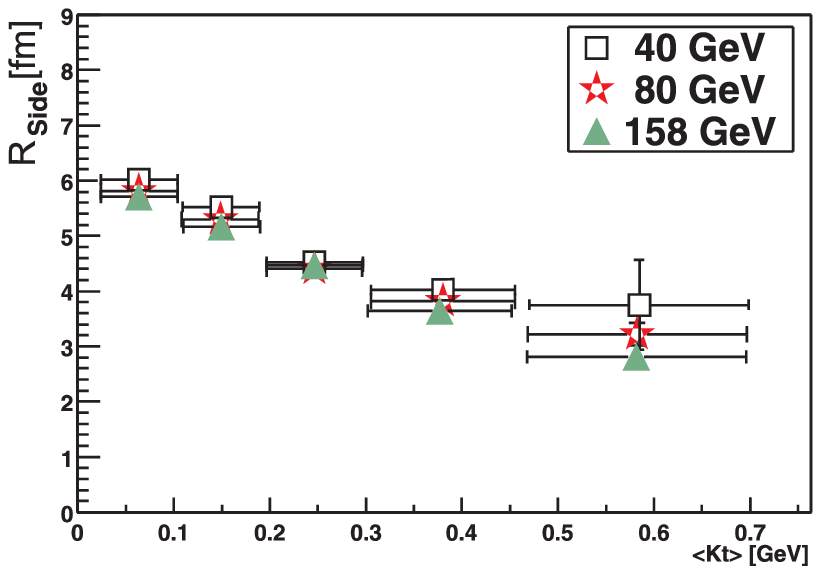,width=7.0cm}}
 \parbox{7.0cm}{
  \epsfig{figure=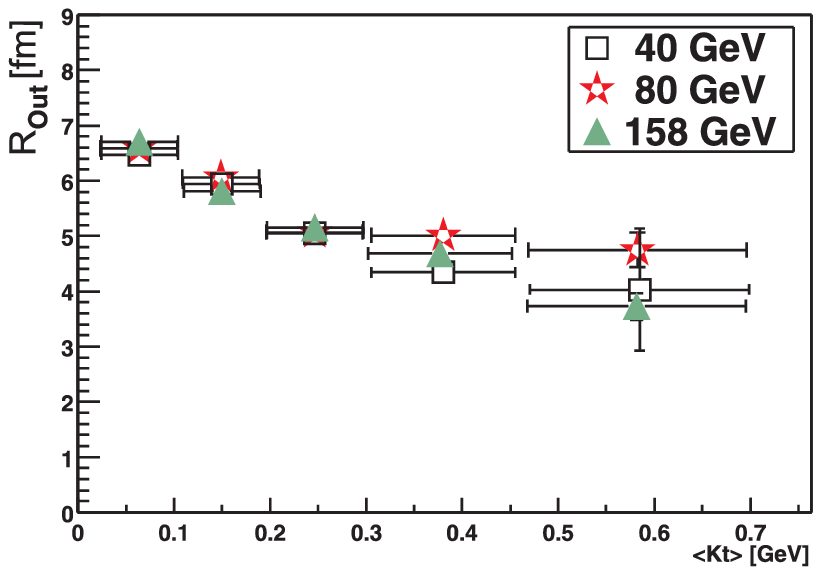,width=7.0cm}}
}
\mbox{
 \parbox{7.0cm}{
  \epsfig{figure=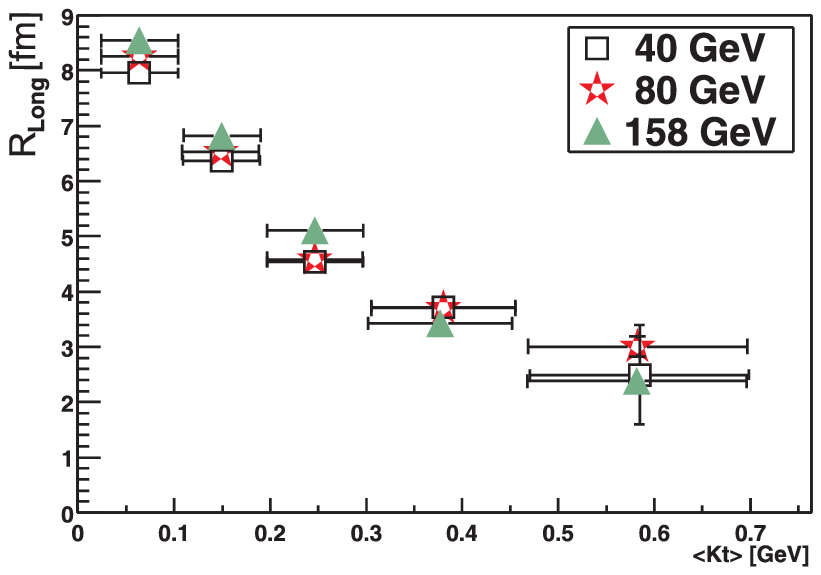,width=7.0cm}}
 \parbox{7.0cm}{
  \epsfig{figure=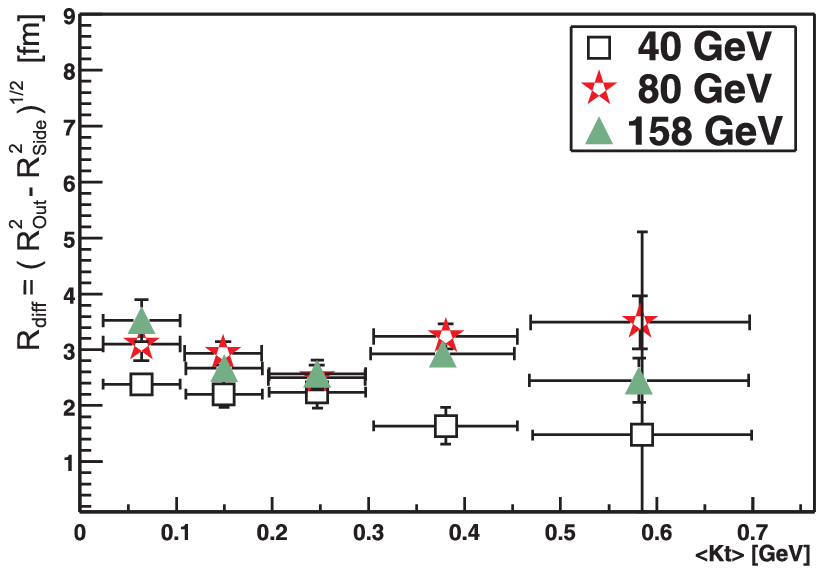,width=7.0cm}}
}
\end{center}
\caption{
Gaussian radius parameters R$_{side}$, R$_{out}$, R$_{long}$ fitted
to the $\pi^-\pi^-$ correlation function evaluated in the longitudinally
comoving frame plotted versus the average transverse momentum K$_T$
of the pair. Rapidity range y$^{\star}\leq$ y $\leq$ y$^{\star}$+0.5 (NA49 preliminary). 
}
\label{fig_hbtradii}
\end{figure}

\section{$\pi\pi$ correlations}\label{correlations}

Correlations of pions with near equal momenta {\bf p$_1$, p$_2$} provide information on the size
and internal dynamics of the fireball source at freezout \cite{wied99}. The analysis
was performed in the longitudinally comoving reference frame, decomposing
the momentum difference {\bf Q = p$_1$ - p$_2$} into long, side, out components. For $\pi^+\pi^-$
pairs the correlation peak at small Q is predominantly due to the Coulomb
attraction and can be well reproduced by a Coulomb wave calculation \cite{siny98}
using the measured effective source size. Correlations of $\pi^-\pi^-$ are described
by a product of parameterised Coulomb repulsion \cite{siny98} and the quantum statistics 
enhancement, fitted with a Gaussian parameterisation. The resulting radius parameters
R$_{side}$, R$_{out}$, R$_{long}$ are plotted in Fig.\ref{fig_hbtradii} versus the
average transverse momentum K$_T$ of the pair. No striking energy dependence is observed.
Within expanding source models the decrease with K$_T$ is a manifestation of the strong
longitudinal (R$_{long}$) and radial (R$_{side}$, R$_{out}$) flow at SPS energies.
Moreover, there is little change in the lifetime $\tau_0\approx$ R$_{long}\sqrt{T/M_T}$
or the emission duration $\Delta\tau^2\approx$ (R$_{out}^2$-R$_{side}^2$)/$\beta_T^2$
of the source. The small value of $\Delta\tau$ does not indicate a soft point of the
matter equation of state of the kind discussed in ref. \cite{hu95} near the onset of
deconfinement. 

\begin{figure}[hbt]
\begin{center}
\mbox{
 \parbox{8.0cm}{
  \epsfig{bbllx=35bp,bblly=60bp,bburx=550bp,bbury=610bp,angle=-90.,
   file=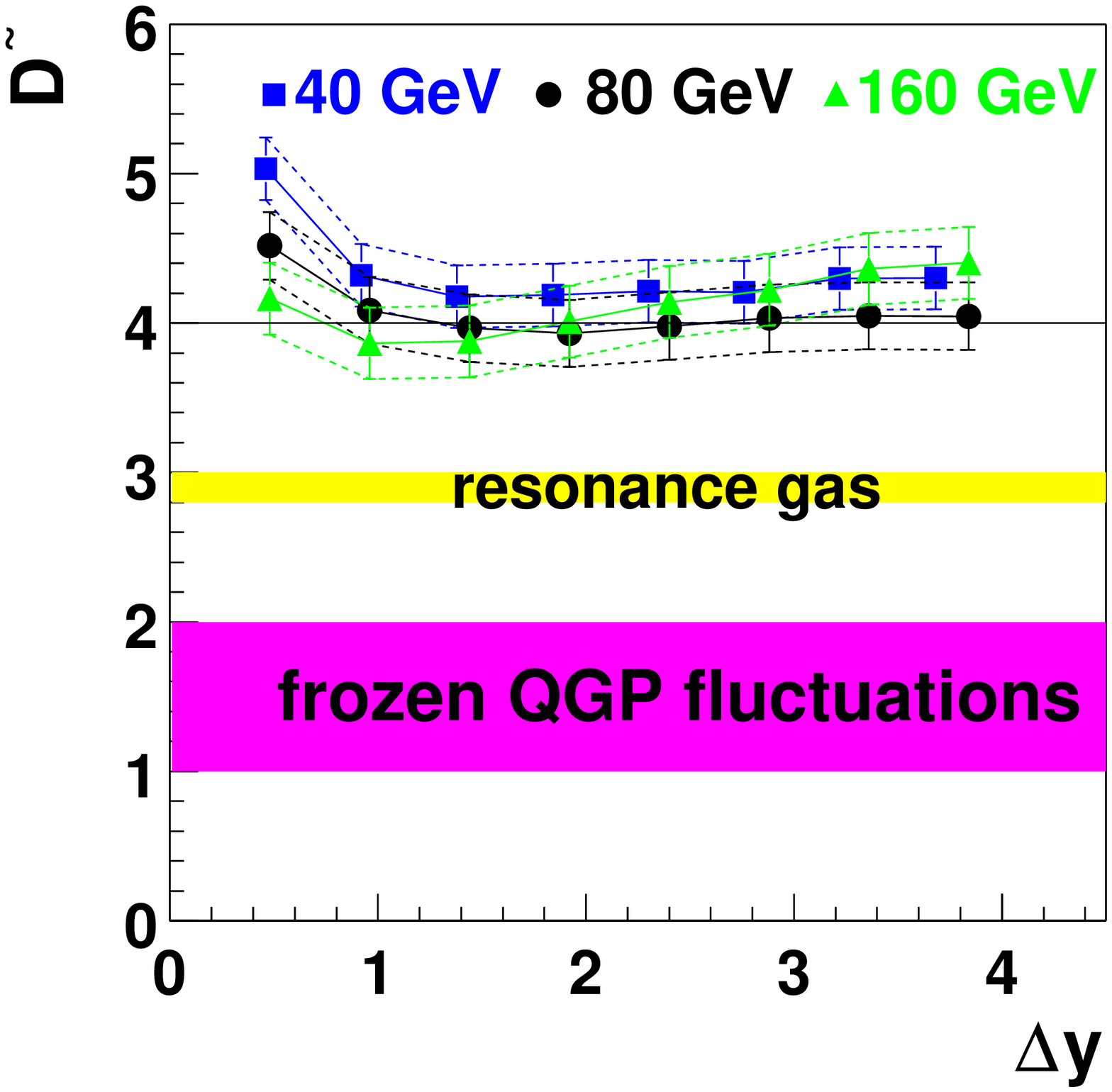,width=8cm}}
 \parbox{8.0cm}{
  \epsfig{bbllx=35bp,bblly=35bp,bburx=570bp,bbury=695bp,angle=-90.,
   file=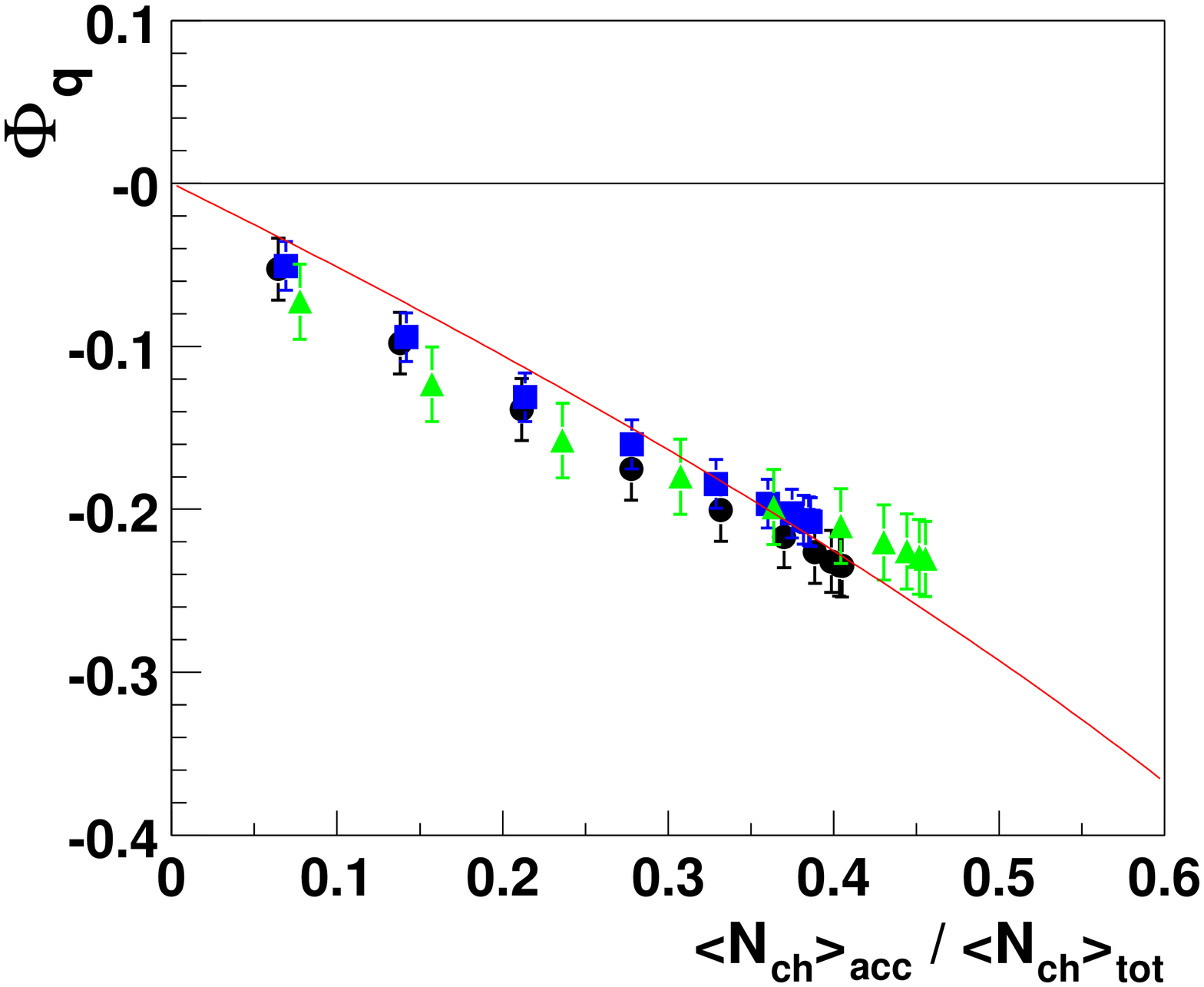,width=8cm}}
}
\end{center}
\caption{
Measures of event-to-event charge fluctuations in central Pb+Pb collisions. Left:
$\tilde{D}$ versus the size of the rapidity window $\Delta$y. Right: $\Phi_{q}$
versus the ratio $\langle N_{ch} \rangle / \langle N_{ch} \rangle _{tot}$ of the 
multiplicity in the acceptance window and the total multiplicity in the events; the curve 
shows the prediction for independent particle emission plus global charge 
conservation (NA49 preliminary).
}
\label{fig_cha_fluct}
\end{figure}

\section{Event-to-event charge fluctuations}\label{fluctuations}

Recently it was proposed that event--to--event fluctuations of the charge ratio
R=N$_+$/N$_-$ or the net charge Q=N$_+$--N$_-$ may be sensitive to deconfinement
in the early stage of nucleus--nucleus collisions \cite{jeon00,asak00}. The 
smaller charge quanta in a partonic phase are expected to result in a 
reduction of such fluctuations.

The fluctuations of the charge ratio were investigated via the measure $\tilde{D}$
\cite{koch00} which is corrected for the residual net charge in the considered
rapidity interval $\Delta$y as well as for global charge conservation. The preliminary
NA49 results are shown in Fig.\ref{fig_cha_fluct} left and are found to be close to 
the expectation for independent particle emission plus global charge conservation 
($\tilde{D}\approx$ 4) and do not change significantly with energy. No evidence is seen 
for the reduction predicted for a resonance gas nor for the large decrease expected 
for a QGP phase. It is, of course, not clear whether reduced fluctuations in the QGP will
persist through the hadronisation, rescattering and resonance decay stages.

The quantity $\Phi_{q}$ was proposed in ref. \cite{gazd99} for studying fluctuations
of the net charge. It is independent of the number of superimposed particle sources,
has value zero for independent particle emission and -1 for local charge conservation.
Preliminary NA49 measurements are plotted in Fig.\ref{fig_cha_fluct} right versus the 
ratio $\langle N_{ch} \rangle / \langle N_{ch} \rangle _{tot}$ of the multiplicity 
in the acceptance window and the total multiplicity in the events. Again no significant 
energy dependence is observed and the results are close to the prediction for 
independent particle emission plus global charge conservation
$\Phi_q^{cc} = \sqrt{1-\langle N_{ch} \rangle / \langle N_{ch} \rangle _{tot}} - 1$ 
(ref. \cite{zara01}, line in Fig.\ref{fig_cha_fluct}).

\begin{figure}[hbt]
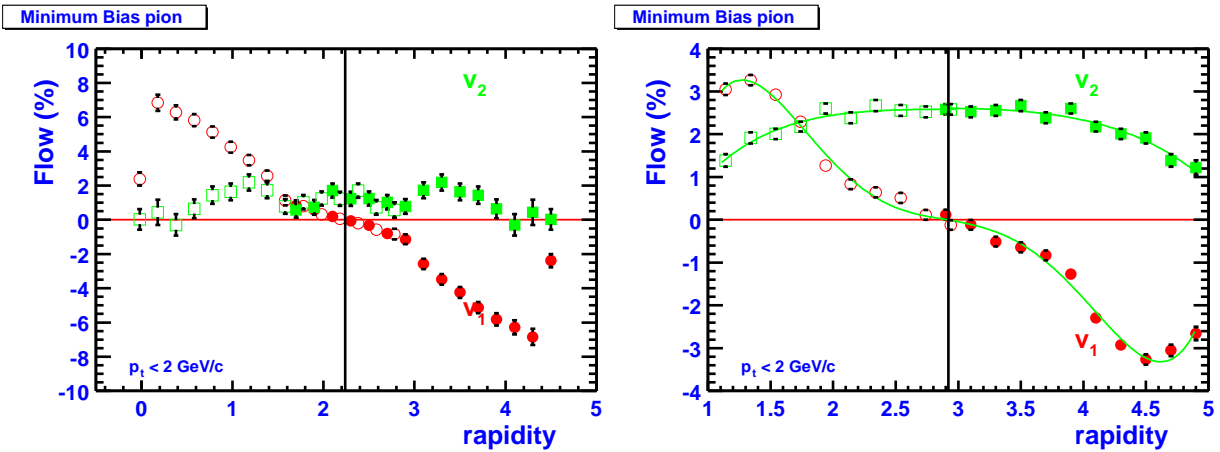

\begin{center}
\mbox{
 \parbox{8.0cm}{
  \epsfig{figure=v_y_minbias_40.epsi,width=8.0cm}}
 \parbox{8.0cm}{
  \epsfig{figure=v_y_minbias_158.epsi,width=8.0cm}}
}
\end{center}
\caption{
Fourier coefficients $v_1$ (directed flow) and $v_2$ (elliptic flow) of the 
azimuthal distribution of pions in min.bias Pb+Pb collisions versus the 
rapidity at 40 (left) and 158 (right) A$\cdot$GeV beam energy (NA49 preliminary).
}
\label{fig_flow}
\end{figure}

\section{Anisotropic flow}\label{flow}

Anisotropic flow in non-central collisions is sensitive to pressure in the
early stage of the reaction, which can transform the initial space anisotropy
of the reaction zone into an azimuthal anisotropy of the momentum distribution
of the observed particles. The onset of deconfinement might result in a
minimum of this effect \cite{hu95}. Anisotropic flow is quantified by the
Fourier coefficients $v_n$ of the distribution of particle azimuthal angles $\Phi$
with respect to the reaction plane $\Psi$ \cite{posk98}:

\centerline{$v_n = \langle cos( n (\Phi_i - \Psi_n) )\rangle
    = \sqrt{2} \langle sin (n \cdot \Phi_i) \cdot sin (n \cdot \Psi_n) \rangle$}

Corrections for reaction plane resolution, nonuniform azimuthal acceptance
 and momentum conservation were applied.
The results for $v_1$ (directed flow) and $v_2$ (elliptic flow) for pions are
plotted versus rapidity in Fig.\ref{fig_flow}. The values for $v_1$ decrease from
40 (left) to 158 (right) A$\cdot$GeV by a factor of 2, whereas $v_2$ shows a
slight increase.

\section{Conclusion}\label{conclusion}

The study of central Pb+Pb collisions in NA49 at 40, 80 and 158 A$\cdot$GeV
led to the following conclusions:
\begin{itemize}
  \item the produced number of pions per participant in Pb+Pb collisions
        changes from suppression with repsect to p+p reactions
        to enhancement in the SPS energy range
  \item the fraction of produced particles containing $s$ or $\bar{s}$ quarks
        passes through a maximum at low SPS energies
  \item strangeness production starts to be undersaturated with respect to
        statistical equilibrium at SPS energies at a level consistent
        with the deconfinement hypothesis
  \item no unusual features are found in the evolution of other characteristics
        of the produced hadron system
\end{itemize}
NA49 will close the data gap between existing measurements at the AGS and the SPS 
with runs at 20 and 30 A$\cdot$GeV in 2002.

\vspace{1.0cm}
\noindent
{\bf Acknowledgements}

This work was supported by the Director, Office of Energy Research, 
Division of Nuclear Physics of the Office of High Energy and Nuclear Physics 
of the US Department of Energy (DE-ACO3-76SFOOO98 and DE-FG02-91ER40609), 
the US National Science Foundation, 
the Bundesministerium fur Bildung und Forschung, Germany, 
the Alexander von Humboldt Foundation, 
the UK Engineering and Physical Sciences Research Council, 
the Polish State Committee for Scientific Research (5 P03B 13820 and 2 P03B 02418), 
the Hungarian Scientific Research Foundation (T14920 and T23790),
the EC Marie Curie Foundation,
and the Polish-German Foundation.

\end{document}